\documentclass[aps,prl,10pt,twocolumn,showpacs,superscriptaddress]{revtex4-1}

\usepackage{latexsym}
\usepackage{graphics,epstopdf}
\usepackage{newlfont}
\usepackage{amssymb}
\usepackage{amsfonts}
\usepackage{amsmath}
\usepackage{amsthm}
\usepackage{graphicx}
\usepackage{epsfig}
\usepackage{color}
\usepackage{bm}
\usepackage{times}

\newcommand{\ket}[1]{|{#1}\rangle}
\newcommand{\bra}[1]{\langle{#1}|}

\begin{document}
%\title{Transition Probability as a Weak Value with Remote Pre and Post selections}
%\title{Measurement of Weak Values with Remote Pre and Post selections and Shared Entanglement}
\title{Weak Values with Remote Postselection and Shared Entanglement}

\author{Arun Kumar Pati}
\author{Uttam Singh}

\affiliation{Quantum Information and Computation Group\\
Harish-Chandra Research Institute, Chhatnag Road, Jhunsi, 
Allahabad 211 019, India}

\date{\today}
\begin{abstract}
We propose a new protocol for the weak measurement of any observable with remote pre and postselections.
We show that if two parties share a pure entangled state, then by using local operations and classical
communication they can preselect and postselect at
distant locations leading to the weak value of an observable as a shift in the pointer of the apparatus
at one location in the process of the weak measurement. This can be achieved with either sharing of a pure
maximally or non-maximally entangled state. We generalize the protocol for realizing the weak value of
any observable with remote pre and postselection of mixed states. Finally, we show how the weak value is
modified in the remote pre and postselection setting if Alice and Bob share a mixed entangled state.
\end{abstract}

\maketitle

\emph{Introduction.}--
The weak measurement has attracted a lot of attention since its inception. The weak measurement allows us to 
obtain information without causing a large disturbance to the quantum system. By bringing the notion of 
time-symmetric feature in quantum formalism \cite{ABL,benni}, it is possible to have evolution from the past
to the future as well as from the future to the past \cite{LV}. Remarkably, it has been shown that if a quantum system
is preselected in the state $\ket{\psi_i}$ and postselected in the state $\ket{\psi_f}$, then the weak
measurement of an operator results in the so called weak value \cite{aav,ab,av} which can be a complex number.
Indeed, both the real and the imaginary parts of the weak value have physical meaning \cite{rj}. Also the
standard expectation value of any observable can be decomposed in to the average of weak values of the
observable \cite{AV90,JT07,HK11,DJ12}. The notion of weak measurement has been generalized beyond its
original formulation \cite{joh,sw1,ADL,sw2,nori,shik}. After several experimental verifications, it has been realized
that the weak value is indeed useful in a broad area of science and technology. For example, the weak
measurements are used in interrogating quantum systems in a coherent manner \cite{toll,cho}, protecting
quantum entanglement from decoherence \cite{kim}, amplifying small experimental effects \cite{wise,mir},
resolving Hardy's paradox \cite{js}, analyzing tunneling time \cite{AMS951, AMS952}, modifying the decay
law \cite{pdav} and in Bohmian trajectories \cite{bohmian1,bohmian2,bohmian3,bohmian4}. Often it is useful
to express certain quantum mechanical quantity of interest in terms of the weak value in order to understand
them better. As an example, the Pancharatnam geometric phase is found to be associated with the phase of a
complex valued weak value arising in a particular type of the weak measurement \cite{eric} and this representation
is then used to study the nontransitive nature of the relative phase in quantum mechanics. Similarly, in quantum
metrology the phase sensitivity of a quantum measurement is given by the variance of the imaginary parts of
the weak values of the generators over the different measurement outcomes \cite{hof}. In fact, the possibility
of using the wavefunction as a weak value, led Lundeen {\it et al} to measure the wavefunction of single
photon directly \cite{jsl1,jsl2}. 

In quantum information and quantum communication, entanglement has played a pivotal role starting from quantum 
teleportation \cite{teleport}, quantum cryptography \cite{BB84,crypto} to remote state preparation \cite{akp,ben}
and many more. In this letter, we raise an important question: can entanglement help in measuring the weak value
of an observable with parties located at distant labs and preselecting and postselecting at remote locations?
We will show that if Alice and Bob share a pure entangled state, and if Alice performs a weak measurement of
some observable, then by using local operations and classical communication Bob can postselect his particle
which has an effect on the pointer of the apparatus
at Alice's lab leading to a shift by an amount equal to the weak value of the observable. {\em Thus, by
postselecting a remote particle that has not interacted with the initial system and apparatus, gives rise
to a shift in the pointer of the measuring apparatus at a remote location}. This is a completely counter
intuitive effect with no classical analog. This we will prove for general observable. We give a proof for
qubit system, but the result can be generalized to higher dimensional Hilbert spaces.  More generally,
we will show that the weak value of any observable for pre and postselected density operator can be realized
with remote pre and postselections if Alice and Bob share a pure entangled state. Thus, our protocol provides
access to directly measure the weak value of any observable at one location having control over postselection
at a distant location. Furthermore, we show that if Alice and Bob share a mixed entangled state, then
remote postselection leads to a quantity that is an admixture of the weak value and the average value.
We argue that pure state entanglement is sufficient for realizing a weak value with remote pre and
postselections.

\emph{Weak value with remote postselection.}--
Let us imagine that Alice and Bob are at two different locations and they share a maximally entangled
state $\ket{\psi^-}_{23}$. Alice has a quantum system prepared in the state
$\ket{\psi_i}_1$. She would like to perform a weak measurement on her particle and measure a weak
value of an observable $A$. Now consider the composite system in the initial state given by
\begin{align}
 \ket{\Psi_{in}}_{123}=\ket{\psi_i}_1\otimes\ket{\psi^-}_{23},
\end{align}
where $\ket{\psi_i}=\alpha\ket{0}+\beta\ket{1}$ is initial state of the qubit and $\ket{\psi^-}$ is the
standard singlet state which is also one of the elements of the Bell basis. One can imagine that this is the preselected state
for the tripartite system, where particles $12$ and $3$ are at remote locations. Now Alice chooses an observable to make a weak
measurement. 
%The observable is given by a projection operator 
%\begin{align}
% A= O \otimes I.
%\ket{\psi^-}_{23}\bra{\psi^-}.
%\end{align}
The weak measurement can be realized using the interaction between the system and the 
measurement apparatus which is governed by the interaction Hamiltonian
\begin{align}
 H_{int}= g \delta(t-t_0)A \otimes P_a,
\end{align}
where $g$ is the strength of the interaction and it is assumed that the
interaction is instantaneous. For sufficiently large $g$, this becomes the usual von Neumann measurement \cite{von}.
%which lasts over a duration $(t-t_0)$ (say).
The operator $A$
%_{123}=O_1\otimes I_{23}$, 
is a Hermitian observable of the system to be measured and
$P$ is an observable
of the apparatus with a conjugate observable $Q$ which works as a pointer. 
In the spectral decomposition, we have $A=\sum_n a_n\ket{a_n}\bra{a_n}$.
Let the initial state of the apparatus be $\ket{\Phi}_a$.
Under the action of the interaction Hamiltonian, the system and apparatus evolve as
\begin{align}
 &\ket{\psi_i}_1 \otimes \ket{\Phi}_a \otimes \ket{\psi^-}_{23}\nonumber\\
&~~~~~~~~~~~~ \rightarrow e^{-\frac{i}{\hbar}g A \otimes P_a} \ket{\psi_i}_1 \otimes \ket{\Phi}_a  \otimes\ket{\psi^-}_{23}.
\end{align}
After the system and apparatus interaction, Alice and Bob perform local projections  using the normal strong measurement. 
The remote postselected state for $12$ and $3$ is given by 
\begin{align}
 \ket{\Psi_{fin}}=\ket{\psi^-}_{12}\otimes\ket{\psi_f}_3.
\end{align}
This can be achieved by local operations and classical communication. For example, Alice makes a Bell state 
measurement on particles $12$. Even though she can get four possible outcomes, she is interested only for the 
outcome $\ket{\psi^-}$ \cite{note}. She can communicate her measurement outcome over a classical channel
to Bob, who can perform a projective measurement on the
particle $3$ in the state $\ket{\psi_f}$. If Bob succeeds in the postselection, he communicates over
a classical channel to Alice.
This we call postselection at remote location. 
After the weak measurement, the shift in the
pointer of the apparatus is given by the weak value of the observable
\begin{align}
\label{totweak}
 {_f\langle A \rangle_i}^w = \frac{\bra{\Psi_{fin}} A \ket{\Psi_{in}}}{\bra{\Psi_{fin}}{\Psi_{in}\rangle}}.
\end{align}
Note that the weak value of $A$ in the tripartite pre and postselected states is actually the weak value 
for $A \otimes  I \otimes I$. But for simplicity we write this as ${_f\langle A \rangle_i}^w$. 
We will show that this weak value is same as the weak value of $A$ as if Alice has preselected in the state $\ket{\psi_i}$,
postselected in the state $\ket{\psi_f}$ and performed the weak measurement on her system alone.
Let us calculate the transition amplitude $\bra{\Psi_{fin}} A \ket{\Psi_{in}}$ 
in the above equation for the observable $A$.
This is given by 
\begin{align}
&\bra{\Psi_{fin}}A\ket{\Psi_{in}} \nonumber\\
&= \big[{_{12}\bra{\psi^-}}\otimes{_3\bra{\psi_f}}\big]\big[\sum_na_n\ket{a_n}_1\bra{a_n}\otimes I_{23}
\big]\big[\ket{\psi_i}_1\otimes\ket{\psi^-}_{23}\big]\nonumber\\
&=\sum_na_n\big[{_{12}\bra{\psi^-}}{_3\bra{\psi_f}}\big]\cdot
\big[\ket{a_n}_1\otimes\ket{\psi^-}_{23}\big]{_1\bra{a_n}}\psi_i\rangle_1\nonumber\\
&=\sum_na_n\big[{_{12}\bra{\psi^-}}{_3\bra{\psi_f}}\big]\cdot\frac{1}{2}\Big[\sum_{i=1}^{4}\ket{B_i}_{12}\otimes
U_i\ket{a_n}_3\Big] {_1\langle{a_n}|}\psi_i\rangle_1\nonumber\\
&=-\frac{1}{2} \sum_na_n \langle{\psi_f}{|a_n\rangle} \langle{a_n}|{\psi_i\rangle}
%&=-\frac{1}{2} {_3\langle{\psi_f}}\Big[|\sum_na_n|a_n\rangle_3
%{_1\langle{a_n}|}\Big]|\psi_i\rangle_1
 =-\frac{1}{2}\langle{\psi_f}|A|\psi_i\rangle.
\end{align}
In the above we have used the fact that $\ket{a_n}_1\otimes\ket{\psi^-}_{23}$ can
be written as
\begin{align}
\label{deco}
\ket{a_n}_1\otimes\ket{\psi^-}_{23}=  \frac{1}{2} \sum_{i=1}^{4}\ket{B_i}_{12} \otimes U_i\ket{a_n}_3,
\end{align}
where $\ket{B_1}=\ket{\phi^+}=\frac{1}{\sqrt{2}}(\ket{00}+\ket{11}),
\ket{B_2}=\ket{\phi^-}= 
\frac{1}{\sqrt{2}}(\ket{00}-\ket{11}),
\ket{B_3}=\ket{\psi^+}=\frac{1}{\sqrt{2}}(\ket{01}+\ket{10}),$
and $ \ket{B_4}=\ket{\psi^-}=\frac{1}{\sqrt{2}}(\ket{01}-\ket{10})$.
The $U_i$'s are the unitary matrices, i.e., $U_1 = - \sigma_z \sigma_x, ~ 
U_2=\sigma_x,~ U_3=-\sigma_z,$ and $U_4= -I$. 
Now, the overlap between the pre and postselected states is given as
$\bra{\Psi_{fin}}\Psi_{in}\rangle = -\frac{1}{2}\langle{\psi_f}|\psi_i\rangle$.
Therefore, we have the desired weak value ${_f\langle A \rangle_i}^w=
\frac{\langle{\psi_f}|A|\psi_i\rangle}{\langle{\psi_f}|\psi_i\rangle}$ which will be seen as a shift of 
the pointer of the apparatus at Alice's lab.

Thus, by sharing an EPR pair and with remotely preselected and postselected quantum systems
Alice can measure weak value of an observable 
of the quantum system. This also shows that the apparatus state is indeed affected by the remote postselection. 
One may wonder, is it that the sharing of maximally entangled state did some magic? In fact, one can show that if 
Alice and Bob share a  general pure entangled state then with a suitable choice of the remote preselected and the
postselected states, the 
pointer of the apparatus is shifted by the weak value of the observable.
%transition probability is indeed a weak value.

Let us consider the situation where Alice and Bob share a non-maximally pure entangled state. Now the preselected state
for Alice and Bob is given by
\begin{align}
 \ket{\Psi_{in}}_{123}=\ket{\psi_i}_1\otimes\ket{\phi_{n}^+}_{23},
\end{align}
where $\ket{\phi_n^+}=\frac{1}{\sqrt{1+|n|^2}}(\ket{00}+n\ket{11})$ is a 
general non-maximally pure entangled state shared between Alice and Bob \cite{pankaj,pati}.
Let Alice makes a weak measurement of the observable by attaching an apparatus to the 
system qubit. Under the action of the interaction Hamiltonian, the system and the apparatus evolve as
\begin{align}
 &\ket{\psi_i}_1 \otimes \ket{\Phi}_a \otimes  \ket{\phi_{n}^+}_{23}\nonumber\\
&~~~~~~~~~~~~ \rightarrow e^{-\frac{i}{\hbar}g A \otimes P_a} \ket{\psi_i}_1 \otimes \ket{\Phi}_a  \otimes\ket{\phi_{n}^+}_{23}.
\end{align}
After the system and apparatus interaction, Alice and Bob can postselect their respective quantum systems in the state given by
\begin{align}
 \ket{\Psi_{fin}}=\ket{\phi_n^-}_{12}\otimes(\sigma_z\ket{\psi_f}_3),
\end{align}
using the normal strong measurements (local operations) and classical communication.
Now, let us calculate the transition amplitude for the observable $A$ which is given by 
\begin{align}
\label{num}
&\bra{\Psi_{fin}}A \ket{\Psi_{in}}\nonumber\\
&= \big[{_{12}\bra{\phi_n^-}}\otimes({_3\bra{\psi_f}}\sigma_z)\big]\nonumber\\
&~~~~~~~~~\cdot\big[\sum_ma_m\ket{a_m}{_1\bra{a_m}}\otimes I_{23} \big]\big[\ket{\psi_i}_1\otimes\ket{\phi_n^+}_{23}\big]\nonumber\\
&=\sum_ma_m\big[{_{12}\bra{\phi_n^-}}\otimes({_3\bra{\psi_f}}\sigma_z)\big]\cdot
\big[\ket{a_m}_1\otimes\ket{\phi_n^+}_{23}\big]{_1\bra{a_m}}\psi_i\rangle_1\nonumber
\end{align}
\begin{align}
&=\sum_ma_m\big[{_{12}\bra{\phi_n^-}}\otimes({_3\bra{\psi_f}}\sigma_z)\big]\nonumber\\
&~~~~~~~~\cdot N^2\Big[\sum_{i=1}^{4}\ket{\tilde{B}_i}_{12}\otimes \ket{a_m^{(i)}}_3\Big]{_1\langle{a_m}}|\psi_i\rangle_1\nonumber\\
&=\sum_ma_m nN^2{_3\langle{\psi_f}}|a_m\rangle_3{_1\langle{a_m}}|\psi_i\rangle_1\nonumber\\ 
&=nN^2\langle{\psi_f}|A|\psi_i\rangle.
\end{align}
Here, in the third line we have used the fact that \cite{pankaj,pati}
\begin{align}
\label{tele}
%\ket{\Psi_{in}}_{123} =
\ket{a_m}_1\otimes\ket{\phi_{n}^+}_{23}
%&=N[\ket{00}_{12}\alpha\ket{0}_3+\ket{01}_{12}\alpha\ket{1}_3+\ket{10}_{12}\beta\ket{0}_3+\ket{11}_{12}\beta\ket{1}_3]\nonumber\\
%&=N^2[\ket{\phi_n^+}_{12}(\alpha\ket{0}+\beta|n|^2\ket{1})_3+n\ket{\phi_n^-}_{12}(\alpha\ket{0}-\beta\ket{1})_3\nonumber\\
%&+n\ket{\psi_n^+}_{12}(\beta\ket{0}+\alpha\ket{1})_3+\ket{\psi_n^-}_{12}(-\beta\ket{0}+\alpha|n|^2\ket{1})_3]\nonumber\\
&= N^2\sum_{i=1}^{4}\ket{\tilde{B}_i}_{12} \otimes \ket{a_m^{(i)}}_3,
\end{align}
where $N=\frac{1}{\sqrt{1+|n|^2}}$, $\ket{a_m} = (c_m \ket{0} + d_m \ket{1})$, $\ket{\tilde{B}_1}=\ket{\phi_n^+}=N (\ket{00}+n\ket{11}),
\ket{\tilde{B}_2}=\ket{\phi_n^-}=N (n^*\ket{00}-\ket{11}),  \ket{\tilde{B}_3}=\ket{\psi_n^+}=N (\ket{01}+n^*\ket{10}),$
and $\ket{\tilde{B}_4}=\ket{\psi_n^-}= N (n\ket{01}-\ket{10})$.
The states $\ket{a_m^{(i)}} $ are the non-normalized states and are given by 
$\ket{a_m^{(1)}}= (c_m \ket{0}+ d_m |n|^2\ket{1}), \ket{a_m^{(2)}} = n (c_m \ket{0}- d_m \ket{1}), 
\ket{a_m^{(3)}}= n (d_m \ket{0}+ c_m \ket{1}),$ and $\ket{a_m^{(4)}} = (-d_m \ket{0}+ c_m |n|^2\ket{1})$.
The overlap between the pre and postselected states is given by $\bra{\Psi_{fin}} \Psi_{in} \rangle = 
nN^2 \langle{\psi_f}|\psi_i\rangle$.
%\begin{align}
%\label{over}
% \bra{\psi_{fin}}\psi_{in}\rangle&=\big[{_{12}\bra{\phi_n^-}}\otimes({_3\bra{\psi_f}}\sigma_z)\big]
% \big[\ket{\psi_i}_1\otimes\ket{\psi^-}_{23}\big]\nonumber\\
% &=\big[{_{12}\bra{\phi_n^-}}\otimes(_3\bra{\psi_f}\sigma_z)\big]n|N|^2
% \sum_{j=1}^{4}\ket{\tilde{B}_j}_{12}\otimes \ket{\psi_i^j}_3\nonumber\\
% &=n|N|^2\langle{\psi_f}|\psi_i\rangle.
%\end{align}
Therefore, the weak value for the observable with remote pre and postselection is given by
 ${_f\langle A\rangle_i}^w=\frac{\langle{\psi_f}|A|\psi_i\rangle}{\langle{\psi_f}|\psi_i\rangle}$.

This shows that even though Bob has postselected his particle in the state $\sigma_z \ket{\psi_f}$ at a remote location,
the effect is as if Alice
has done postselection of her particle in the state $\ket{\psi_f}$. 
%One thing may be worth emphasizing here. 
One may wonder how the preselection and postselection at remote locations have an effect on the pointer
state of the apparatus? It is the quantum entanglement that provides an invisible link between the past and the future.
One can understand this by saying that a part of the contribution to the amplitude of the weak value 
comes from the Hilbert space ${\cal H}_1$ and other part to the amplitude comes from the Hilbert space ${\cal H}_3$.
This is possible because of the existence of the shared entanglement between Alice and Bob.
Thus, the contribution to the weak value of $A$ indeed comes from the past and the future (not only from a future 
time but also from a future space, possibly).

%\emph{Transition probability as weak value with remote postselection}--
As a simple application, one can measure the transition amplitude and transition probability with remote pre and postselections.
To measure transition probability with remote postselection, let us imagine that Alice and Bob are at two different locations and 
they share a maximally entangled state. 
Alice has a quantum system prepared in the state $\ket{\psi}_1$. She would like to 
perform a weak measurement on her particle and measure a weak value so as to find the transition probability 
between two non-orthogonal states $\ket{\psi}$ and $\ket{\phi}$, i.e., $|\langle{\psi}\ket{\phi}|^2$ without causing 
too much disturbance to the quantum system.
From the above expression, we can see that for the special choices $A = \ket{\phi}\bra{\phi}$ and $\ket{\psi_f}=\ket{\psi}$,
we have ${_f\langle A \rangle_i}^w= |\bra{\phi}\psi\rangle|^2$.

%Again for the special choices of $\ket{\psi_i}=\ket{\psi}$, $O=\ket{\phi}\bra{\phi}$ and $\ket{\psi_f}=\ket{\psi}$,
%we have ${_f\langle A \rangle_i}^w=|\bra{\phi}\psi\rangle|^2$,
%which shows that the transition probability can be realized as a weak 
%value using the remotely chosen pre and post-selected ensembles and sharing of entanglement.
%Moreover, our result is robust to the choice 
%of the overlap between the preselected and postselected state. For example, from (\ref{num}) and (\ref{over})
%it follows that one can choose
%even a very small value of $n$, making the overlap also small, but the weak value is independent of that.

\emph{ Weak value for mixed states with remote postselection.}--
The protocol for the weak measurement with remote pre and postselections can be generalized to mixed quantum states.
Let Alice and Bob share a pure maximally entangled state. Let the initial preselected state is given by
\begin{align}
 \chi_{in} =\rho^i_1 \otimes\ket{\psi^-}_{23}\bra{\psi^-},
\end{align}
where $\rho^i_1=\sum_lp_l\ket{l}_1\bra{l}$ is the density matrix for Alice's qubit.
Now Alice carries out a weak measurement by coupling the system observable to an apparatus observable 
with the interaction Hamiltonian.
%\begin{align}
% A=O_1\otimes I_{23},
%\end{align}
%where $O=\sum_na_n\ket{a_n}\bra{a_n}$.
After the interaction between the system and the apparatus, Alice and Bob can postselect (via LOCC)
the whole system in the state given by
\begin{align}
 \chi_{fin} =\ket{\psi^-}_{12}\bra{\psi^-}\otimes \rho^f_3,
\end{align}
where $\rho^f=\sum_{j,k}q_{jk}\ket{j}\bra{k}$. After the weak measurement the shift in the
pointer of the apparatus at Alice's lab is given by the weak value of the observable $A$ \cite{DJ12}
\begin{align}
 {_f\langle A\rangle_i}^w = \frac{\mathrm{Tr}_{123}\big[\chi_{fin} A \chi_{in} \big]}
 {\mathrm{Tr}_{123}\big[\chi_{fin} \chi_{in} \big]}.
 %\equiv \frac{N_1}{D_1}.
\end{align}
What we will show is that, this weak value is indeed given by  
\begin{align}
{_f\langle A\rangle_i}^w=\frac{\mathrm{Tr}[\rho^f  A \rho^i]}{\mathrm{Tr}[\rho^f\rho^i]}
\end{align}
as if the preselection and postselection have been done with 
$\rho^i$ and $\rho^f$, respectively in Alice's lab, even though the postselection has been done by Bob at a 
remote location.

To show this explicitly, we calculate $\mathrm{Tr}_{123}\big[\chi_{fin} A \chi_{in} \big]$. This is given by 
\begin{align}
&\mathrm{Tr}_{123}\big[\chi_{fin}A\chi_{in}\big]
%=\mathrm{Tr}_{123}\Big\{\big[\ket{\psi^-}_{12}\bra{\psi^-}\otimes
%\sum_{j,k}q_{jk}\ket{j}_3\bra{k}\big]\nonumber\\
%&\cdot\big[\sum_na_n\ket{a_n}\bra{a_n}\otimes I_{23}\big]\cdot\big[\sum_lp_l\ket{l}_1\bra{l}
%\otimes\ket{\psi^-}_{23}\bra{\psi^-}\big]\Big\}\nonumber\\
=\mathrm{Tr}_{123}\Big\{\sum_{l,j,k,n}p_lq_{jk}a_n\bra{a_n}{l}\rangle\nonumber\\
&\cdot\big[\ket{\psi^-}_{12}\bra{\psi^-}\otimes\ket{j}_3\bra{k} \big]
\big[\ket{a_n}_1\bra{l}\otimes\ket{\psi^-}_{23}\bra{\psi^-}\big]\Big\}\nonumber\\
&=\frac{1}{4}\mathrm{Tr}_{3}\Big\{\sum_{l,j,k,n}p_lq_{jk}a_n\bra{a_n}{l}\rangle \bra{k}{a_n}\rangle \ket{j}_3\bra{l}\Big\} \nonumber\\
%&= \frac{1}{4}\sum_{l,k,n}p_lq_{lk}a_n\bra{k}{a_n}\rangle \bra{a_n}{l}\rangle 
&= \frac{1}{4}\sum_{l,k}p_lq_{lk}\bra{k} A  \ket{l}=\frac{1}{4}\mathrm{Tr}[\rho^f A \rho^i].
\end{align}
%where in the third line, we have used (\ref{deco}).
Now
the denominator is given by $\mathrm{Tr}_{123}\big[\chi_{fin}\chi_{in}\big] = \frac{1}{4}\mathrm{Tr}[\rho^f\rho^i]$.
%\begin{align}
% &\mathrm{Tr}_{123}\big[\chi_{fin}\chi_{in}\big] =\mathrm{Tr}_{123}\Big\{\big[\ket{\psi^-}_{12}\bra{\psi^-}
% \otimes \sum_{j,k}q_{jk}\ket{j}_3\bra{k}\big]\nonumber\\
%&\cdot\big[\sum_lp_l\ket{l}_1\bra{l}\otimes\ket{\psi^-}_{23}\bra{\psi^-}\big]\Big\}\nonumber\\
%&=\frac{1}{4} \sum_{l,j,k}p_lq_{jk}\bra{k}{l}\rangle\delta_{lj}=\frac{1}{4} \sum_{l,k}p_lq_{lk}\bra{k}{l}\rangle\nonumber\\
%&=\frac{1}{4}\mathrm{Tr}[\rho^f\rho^i].
%\end{align}
Therefore, we have the weak value of $A$ for the mixed pre and postselected states as 
\begin{align}
{_f\langle A\rangle_i}^w=\frac{\mathrm{Tr}[\rho_f A \rho^i ]}{\mathrm{Tr}[\rho_f \rho_i ]}.
\end{align}
This shows that the weak measurement of an observable $A$ with remote pre and postselections in the states $\chi_{in}$ and $\chi_{fin}$
is equivalent to the weak measurement of $A$ with pre and postselections in the state $\rho^i$ and $\rho^f$, respectively.
%This can be of practical use when Alice may not have the ability to postselect in the state $\rho^f$. 
Thus, if Alice and
Bob share a pure maximally entangled
state and Bob can postselect in $\rho^f$, Alice's apparatus will be shifted by an amount equal to the weak value of the 
observable.

Our result shows that by sharing a pure entangled state and suitably choosing
the remote pre and postselected ensemble via LOCC one can also measure the
weak value in distant laboratory paradigm. This can possibly have multitude of
ramifications in quantum mechanics, quantum information, quantum metrology and other areas of physics. For 
example, one can directly measure the expectation value of any observable without causing the 
collapse of the quantum state and having control over a remote location. By gently or weakly measuring
the quantum system in a way that is prescribed by the formalism of the weak measurement and
following our protocol of preparing the pre and postselected ensembles
at remote locations, one can read out the expectation value from the shift of the meter of the apparatus.

%*****************************************************************************************************************

\emph{ Weak values with remote postselection and shared mixed entangled state--}
Now, we raise a question that if Alice and Bob share a mixed entangled state, can remote postselection at
Bob's lab still yield a weak value for Alice in her lab? We will show that in this case the weak value
is contaminated by an admixture of the average value of the observable.
Let Alice and Bob share an arbitrary bipartite quantum state $\xi_{23}$ which can be a noisy entangled state.
Let the composite system of Alice
and Bob be in the initial preselected state as given by
\begin{align}
 \rho^{in}_{123}=\rho^i_1\otimes\xi_{23},
\end{align}
where $\rho^i$ is a general qubit density matrix. Now Alice performs a weak measurement of
observable $A$ on her system $1$, conditioned on the postselections done by Alice and Bob (via LOCC) at remote locations in the state
\begin{align}
\rho^{fin}_{123}=\ket{\psi^-}_{12}\bra{\psi^-}\otimes\rho^f_{3},
\end{align}
where $\rho^f = \sum_kb_k\ket{b_k}\bra{b_k}$. As a result of the weak measurement the shift in the
pointer of the apparatus is given by the weak value of the observable
\begin{align}
\label{wval}
 {_f\langle A \rangle_i}^w = \frac{\mathrm{Tr}_{123}[\rho^{fin} A \rho^{in}]}{\mathrm{Tr}_{123}[\rho^{fin} \rho^{in}]}.
\end{align}
The numerator of the above equation is given by 
\begin{align}
\label{nume}
&\mathrm{Tr}_{123}[\rho^{fin} A \rho^{in}]\nonumber\\
&= \mathrm{Tr}_{123}\big[ \{ \ket{\psi^-}_{12}\bra{\psi^-}\otimes\rho^f_{3}\}
\{(A\rho^i)_1\otimes\xi_{23}\} \big]\nonumber\\
&= \frac{1}{4}\mathrm{Tr}_{123}\sum_{m,n}\big[  V_m\rho^f_1V^\dagger_n(A\rho^i)_1\otimes\ket{B_m}_{23}\bra{B_n}\xi_{23}
 \big]\nonumber\\
 &=\frac{1}{4} \sum_{m,n} \mathrm{Tr} [V_m\rho^fV^\dagger_n A\rho^i] \bra{B_n}\xi\ket{B_m},
\end{align}
where $V_m$'s are appropriate unitaries (in terms of the Pauli matrices).
%In the above we have used $\ket{\psi^-}_{12}\otimes\ket{\psi}_3 = \sum_{m}V_m\ket{\psi}_1\otimes\ket{\psi^-}_{23}$ to write 
%\begin{align}
%\ket{\psi^-}_{12}&\bra{\psi^-}\otimes\rho_3\nonumber\\
%&=\frac{1}{4}\sum_{m,n=1}^{4}V_m\rho_1V^\dagger_n\otimes\ket{B_m}_{23}\bra{B_n}.
%\end{align}
%The $V_i$s are unitary matrices given by, $V_1=\sigma_z\sigma_x$, $V_2=-\sigma_x$, $V_3=\sigma_z$, and $V_4=-I$.
Similarly, the denominator is given by
\begin{align}
\label{denom}
\mathrm{Tr}_{123}[\rho^{fin} \rho^{in}]=\frac{1}{4} \sum_{m,n} \mathrm{Tr} [V_m\rho^fV^\dagger_n \rho^i] \bra{B_n}\xi\ket{B_m}.
\end{align}
Therefore, the weak value is given by
\begin{align}
\label{genweak}
 {_f\langle A \rangle_i}^w = \frac{\sum_{m,n} \mathrm{Tr} [V_m\rho^fV^\dagger_n A\rho^i] \xi_{nm} }
 { \sum_{m,n} \mathrm{Tr} [V_m\rho^fV^\dagger_n \rho^i] \xi_{nm} },
\end{align}
where $\xi_{nm} = \bra{B_n}\xi\ket{B_m}$ are the matrix elements of the noisy entangled state in the 
Bell basis. Note that this is not equal to the desired weak value that Alice would have measured as a shift of 
the pointer of the apparatus. This shows that if Alice and Bob share a noisy entangled state, then it is
not possible to measure weak value with remote postselection. However, if $\xi_{23}= \ket{B_k}_{23}\bra{B_k}$ is
one of the Bell states, then we have ${_f\langle A \rangle_i}^w = \frac{\mathrm{Tr} [\rho^f A\rho^i]}{\mathrm{Tr} [\rho^f \rho^i]}$.
Thus, sharing of a pure maximally entangled state is
sufficient to reproduce the weak value in remote pre and postselection scenario.
%Also, we can argue that pure entangled state is sufficient to reproduce the weak value.

%This shows that a shared pure maximally entangled state is needed for measurement of weak value with remote postselection.
Now, we will see how the weak value is modified if Alice and Bob share a pseudo-pure entangled state.
Let us consider the initially shared entangled state to be a pseudo pure entangled state as
given by the Werner state
\begin{align}
 \xi_{23}= p \ket{\psi^-}_{23}\bra{\psi^-} + \frac{(1-p)}{4}I_{23}.
\end{align}
Using (\ref{genweak}), the weak value of the observable $A$ is given by
\begin{align}
 {_f\langle A \rangle_i}^w 
 &= \frac{ \frac{(1+3p)}{4}\mathrm{Tr} [\rho^f A\rho^i] + \frac{(1-p)}{4} Q(A)_i^f}
 {\frac{(1+3p)}{4}\mathrm{Tr} [\rho^f\rho^i] + \frac{(1-p)}{4}Q(I)_i^f},
\end{align}
where $Q(A)_i^f = \sum_{m=1}^{3} \mathrm{Tr} [\rho^fV^\dagger_m A\rho^iV_m]$.
By using the fact that $\sum_{m=1}^{4}V_m\rho^fV_m^\dagger = 2I$, we have 
$Q(A)_i^f = 2\mathrm{Tr} [A\rho^i] - \mathrm{Tr} [\rho^fA\rho^i]$ and 
$Q(I)_i^f = 2 - \mathrm{Tr} [\rho^f\rho^i]$.
% and 
%$Q_D = \sum_{m=1}^{3} \mathrm{Tr} [\rho^fV^\dagger_m \rho^iV_m]$. In general, $A\rho^i$ can be written as 
%\begin{align}
% A\rho^i = aI + \vec{b}.\vec{\sigma},
%\end{align}
%where $a = \frac{1}{2}\mathrm{Tr}[A\rho^i]$ and $b_w = \mathrm{Tr}[\sigma_w A\rho^i], (j=x,y,z)$. Now
%\begin{align}
 %Q_N &= \mathrm{Tr} [\rho^f\{ \sigma_x A\rho^i\sigma_x + \sigma_y A\rho^i\sigma_y + \sigma_z A\rho^i\sigma_z\}]\nonumber\\
 %&= \mathrm{Tr} [\rho^f\{ 3aI - b_x\sigma_x - b_y\sigma_y - b_z\sigma_z\}]\nonumber\\
 %&= 2\mathrm{Tr} [A\rho^i] - \mathrm{Tr} [\rho^fA\rho^i].
%\end{align}
%Similarly, $Q_D = 2 - \mathrm{Tr} [\rho^f\rho^i]$.
Therefore, we have the weak value for the observable with sharing of pseudo pure entangled state as 
\begin{align}
 {_f\langle A \rangle_i}^w 
 &= \frac{ p\mathrm{Tr} [\rho^f A\rho^i] + \frac{(1-p)}{2} \mathrm{Tr} [A\rho^i]}
 {p\mathrm{Tr} [\rho^f\rho^i] + \frac{(1-p)}{2} }.
\end{align}
In this case, the shift in the pointer of the apparatus is an admixture of the ideal weak value and the average value of the operator 
in the initial state. We note 
from the above equation that if Alice and Bob share a maximally pure entangled state, i.e., $p = 1$, we can have 
the ideal weak value of $A$ as if Alice has preselected and postselected her system in the states $\rho^i$ and $\rho^f$, respectively. 
Also,we can check that if we demand that the above weak value is equal to the ideal weak value, then $p=1$. 
Thus, Alice will recover the weak value of the observable
with the help of the remote postselection by Bob when they share a pure maximally entangled state. 
%*****************************************************************************************************************

\emph{Conclusion.}-- 
To conclude, we have raised a fundamental question in this paper: can preselected and
postselected quantum states at two remote locations shift the pointer of
the measuring apparatus at one location, so as to give the weak value of an observable?
We have shown that if two parties share a pure entangled state, then at two distant locations they 
can preselect and postselect using LOCC, leading to the weak value of an observable at one location. This is vividly demonstrated 
%by showing that the transition probability between two 
for
%is equal to the weak value of a projection operator
remotely chosen pre and postselected
mixed states with prior sharing of a pure maximally entangled state. 
%As a simple application, we show that the transition probability can be read directly from the shift in the 
%meter reading of the apparatus that is used for the weak measurement. 
This may provide a direct measurement
of the weak value using the weak measurement having a control over a remote location. 
The use of entangled states and classical communication helps 
in preselection and postselection that sends information about the state from the past to the future and
from the future to the past, respectively. 
The contribution to the weak value that appears as a
%The joint operator on the enlarged Hilbert space gives the right 
shift 
in the apparatus state, comes seemingly from the past and from the future. 
It seems that nature plays right conspiracy to realize the weak value as a
shift in the pointer of the apparatus in Alice's lab in the remotely chosen pre and postselected quantum states. 
This shows another usefulness of the pure maximally entangled state. In future, it will be interesting to generalize the protocol
for the continuous variables \cite{cont}.
This proposal can have potential applications in quantum information, measurement, quantum metrology and quantum foundations.
This may open up new avenues of explorations in performing various quantum tasks with weak measurements under LOCC paradigm.
Since quantum entanglement has already been distributed over long distances \cite{longt}, it is possible to
implement our protocol
with the current technology. This may lead to measurement of the weak value
of any observable with remote postselection over long distances.
\vskip 1cm
\noindent
{\em Acknowledgement:}
We would like to thank L. Vaidman and A. D. Lorenzo for useful remarks.


\begin{thebibliography}{999}

\bibitem{ABL} Y. Aharonov, P. G. Bergmann and J. L. Lebowitz, Phys. Rev. {\bf134}, B1410 (1964).

\bibitem{benni} B. Reznik, and Y. Aharonov, Phys. Rev. A {\bf52}, 2538 (1995). 

\bibitem{LV} L. Vaidman, J. Phys. A: Math. Theor. {\bf40}, 3275–3284 (2007). 


\bibitem{aav} Y. Aharonov, D. Z. Albert, and L. Vaidman, Phys. Rev. Lett. {\bf 60}, 1351 (1988).

\bibitem{ab} Y. Aharonov and A. Botero, Phys. Rev. A {\bf72}, 052111 (2005).

\bibitem{av} Y. Aharonov and L. Vaidman, Lect. Notes Phys. {\bf734}, 399 (2008).

\bibitem{rj} R. Jozsa, Phys. Rev. A {\bf 76}, 044103 (2007).

\bibitem{AV90} Aharonov and L. Vaidman, Phys. Rev. A {\bf41}, 11 (1990).

\bibitem{JT07} J. Tollaksen, J. Phys. A: Math. Theor. {\bf40}, 9033
(2007).

\bibitem{HK11} A. Hosoya and M. Koga, J. Phys. A: Math. Theor. {\bf44}, 415303
(2011).



 
\bibitem{DJ12} J. Dressel and A. N. Jordan, Phys. Rev. A {\bf85}, 012107 (2012).

\bibitem{joh} L. M. Johansen, Phys. Rev. Lett. {\bf 93}, 120402 (2004).

\bibitem{sw1} S. Wu and Y. Li, Phys. Rev. A {\bf83}, 052106 (2011).

\bibitem{ADL} A. D. Lorenzo, Phys. Rev. A {\bf85}, 032106 (2012).

\bibitem{sw2} S. Pang, S. Wu, and Z.B. Chen, Phys. Rev. A {\bf86}, 022112 (2012).

\bibitem{nori} A. G. Kofman, S. Ashhab, and F. Nori, Physics Reports {\bf520}, 43–133 (2012).

\bibitem{shik} Y. Shikano, and A. Hosoya, J. Phys. A: Math. Theor. {\bf43}, 025304 (2010). 

\bibitem{toll} Y. Aharonov, S. Popescu, and J. Tollaksen, Phys. Today {\bf63}, No.1 11, 27 (2010).

\bibitem{cho} A. Cho, Science {\bf333}, 690 (2011).

\bibitem{kim} Y. S. Kim, J. C. Lee, O. Kwon, and Y. H. Kim, Nat. Phys. {\bf8}, 117 (2012).

\bibitem{wise} H. M. Wiseman, Phys. Lett. A {\bf311}, 285 (2003).

\bibitem{mir} R. Mir {\it et al }, New J. Phys. {\bf9}, 287 (2007).

\bibitem{js} J.  S.  Lundeen and A.  M.  Steinberg, Phys. Rev. Lett. {\bf 102}, 020404 (2009).

\bibitem{AMS951} A. M. Steinberg, Phys. Rev. Lett. {\bf74}, 2405 (1995).

\bibitem{AMS952} A. M. Steinberg, Phys. Rev. A {\bf52}, 32 (1995).

\bibitem{pdav} P. C. W. Davies, Phys. Rev. A {\bf79}, 032103 (2009).

\bibitem{bohmian1} C. R. Leavens, Found. Phys. {\bf35}, 469 (2005).

\bibitem{bohmian2} H. M. Wiseman, New J. Phys. {\bf9}, 165 (2007).

\bibitem{bohmian3} S. Kocsis {\it et al}
%B. Braverman, S. Ravets, M. J. Stevens, R. P. Mirin, L. K. Shalm, and A. M. Steinberg, 
Science {\bf332}, 1170 (2011).

\bibitem{bohmian4} B. J. Hiley, J. Phys.: Conference Series, {\bf361}, 012014 (2012).

%\bibitem{dix} P. B. Dixon, D. J. Starling, A. N. Jordan, and J. C. Howell, Phys. Rev. Lett. {\bf 102}, 173601 (2009).

\bibitem{eric} E. Sj{\"o}qvist, Phys. Lett. A {\bf359}, 187 (2006).

\bibitem{hof} H. F. Hofmann, Phys. Rev. A {\bf 83}, 022106 (2011).

\bibitem{jsl1} J. S. Lundeen {\it et al},
%B. Sutherland, A. Patel, C. Stewart, and C. Bamber,
Nature {\bf474}, 188 (2011).

\bibitem{jsl2} J. S. Lundeen and C. Bamber, Phys. Rev. Lett. {\bf108}, 070402 (2012). 

\bibitem{teleport} C. H. Bennett {\it et al},
%G. Brassard, C. Cr$\acute{e}$peau, R. Josza, A. peres, and W. K. Wootters,
Phys. Rev. Lett. {\bf70}, 1895 (1993).

\bibitem{BB84} C. H. Bennett, and G. Brassard, {\it Proceedings of the
IEEE International Conference on Computers, Systems and Signal Processing} (IEEE Computer Society, New York, 1984),
pp. 175–179.

\bibitem{crypto} A. K. Ekert, Phys. Rev. Lett. {\bf 67}, 661 (1991).

\bibitem{akp} A. K. Pati, Phys. Rev. A {\bf 63}, 014302 (2000).

\bibitem{ben} C. H. Bennett {\it et al},
%David P. DiVincenzo, Peter W. Shor, John A. Smolin, Barbara M. Terhal, and William K. Wootters
Phys. Rev. Lett. {\bf87}, 077902 (2001). 

\bibitem{von} J. von Neumann, {\it Mathematical Foundations of Quantum Mechanics},
(Princeton Univ. Press, Princeton, NJ, 1955).

\bibitem{note} The weak value will appear as a shift of the pointer of the apparatus for any one of the four possible outcomes.
For example, after the Bell measurement, Alice communicates two cbits to Bob. Then, Bob can postselect his  quantum system
in the state $U_i \ket{\psi_f}$ and communicate to Alice whether he succeeded in the postselection or not by sending one cbit.
This will result in the weak value of the observable as a shift in the pointer of the apparatus at Alice's lab as if she has
postselected her particle in the state $\ket{\psi_f}$.

\bibitem{pankaj} P. Agrawal and A. K. Pati, Phys. Lett. A {\bf 305}, 12 (2002).

\bibitem{pati} A. K. Pati and P. Agrawal, Phys. Lett. A {\bf 371}, 185 (2007).

\bibitem{cont} S. L. Braunstein and A. K. Pati, {\it Quantum Information with Continuous Variables},
(Kluwer Academic Publishers, The Netherlands, April 2003).

\bibitem{longt} Xiao-Song Ma {\it et al}, Nature {\bf489}, 269 (2012).


%\bibitem{utm} U. Singh and A. Pati, arXiv:1211.0939 (2012).

\end{thebibliography}
\end{document}